\pdfoutput=1
\documentclass[12pt,oneside]{amsart}

\usepackage[top=1in, left=1.25in, right=1.25in, bottom=1in]{geometry}
\usepackage{epsfig}
\usepackage{amsmath}
\usepackage{amssymb}
\usepackage{color}
\usepackage{multirow}
\usepackage{algorithm}
\usepackage{algorithmic}
\usepackage{bbold}
\usepackage{morefloats}
\usepackage{wrapfig}
\usepackage{fancybox}
\usepackage{xcolor}
\usepackage{graphicx}
\usepackage{amssymb}
\usepackage{amsmath}
\usepackage{mathtools}
\usepackage{subfigure}
\usepackage{mdframed}
\usepackage{url}
\usepackage{tikz}
\usetikzlibrary{arrows,shapes,chains,matrix,positioning,scopes,shadows,backgrounds,fit,petri}

\usepackage{hyperref}
\hypersetup{
  colorlinks,
  citecolor=black,
  filecolor=black,
  linkcolor=black,
  urlcolor=black
}

\usepackage{amsfonts}
\usepackage{psfrag}
\usepackage{bbm}
\usepackage{bm}
\usepackage{stmaryrd,dsfont}
\usepackage[latin1]{inputenc}
\usepackage{amsbsy}
\usepackage{listings}

\usepackage{graphicx}
\usepackage{hyperref}
\usepackage{amssymb,amstext,amsmath}
\usepackage{color}
\usepackage{soul}

\usepackage{algorithm}
\usepackage{algorithmic}
\usepackage{bbm}

\newtheorem{conjecture}{Conjecture}

\newcommand{\perm}{\mathrm{perm}}
\newcommand{\Q}{\mathbb{Q}}
\renewcommand{\P}{\mathbb{P}}
\newcommand{\ds}{\mathrm{d}s}
\newcommand{\dt}{\mathrm{d}t}
\newcommand{\dr}{\mathrm{d}r}
\newcommand{\db}{\mathrm{d}b}
\newcommand{\du}{\mathrm{d}u}
\newcommand{\e}{\mathrm{e}}
\newcommand{\cin}{c_{\mathrm{in}}}
\newcommand{\cout}{c_{\mathrm{out}}}
\newcommand{\dmax}{d_{\mathrm{max}}}

\def\E{{\mathbb E}}

\makeatletter
\newcommand\Cshadowbox{\VerbBox\@Cshadowbox}
\def\@Cshadowbox#1{%
  \setbox\@fancybox\hbox{\fbox{#1}}%
  \leavevmode\vbox{%
    \offinterlineskip
    \dimen@=\shadowsize
    \advance\dimen@ .5\fboxrule
    \hbox{\copy\@fancybox\kern.5\fboxrule\lower\shadowsize\hbox{%
      \color{ShadowColor}\vrule \@height\ht\@fancybox \@depth\dp\@fancybox \@width\dimen@}}%
    \vskip\dimexpr-\dimen@+0.5\fboxrule\relax
    \moveright\shadowsize\vbox{%
      \color{ShadowColor}\hrule \@width\wd\@fancybox \@height\dimen@}}}
\makeatother

\newlength{\figurewidth}
\newlength{\smallfigurewidth}

\theoremstyle{plain}

\newtheorem{definition}{Definition}
\newtheorem{example}{Example}

\numberwithin{equation}{section}

\begin{document}

\title[Computation, Phase Transitions, and Community Detection]{Computational Complexity, Phase Transitions, and Message-Passing for Community Detection}
\author{Aur\'elien Decelle, Janina H\"uttel, Alaa Saade, Cristopher Moore}

\begin{center}

\maketitle

\emph{These are notes from the lecture of Cristopher Moore given at the autumn school ``Statistical Physics, Optimization, Inference, and Message-Passing Algorithms'', that took place in Les Houches, France from Monday September 30th, 2013, till Friday October 11th, 2013. The school was organized by Florent Krzakala from UPMC \& ENS Paris, Federico Ricci-Tersenghi from La Sapienza Roma, Lenka Zdeborov\'a from CEA Saclay \& CNRS, and Riccardo Zecchina from Politecnico Torino. Much of this material is covered in Chapters 4, 5, 13, and 14 of~\cite{Moore11}.}
\end{center}
\tableofcontents
\section{Computational complexity}

Computational complexity is a branch of complex systems which has the advantages of being very well defined and presenting a lot of rigorous results. It allows qualitative distinctions between different kinds of ``hardness'' in computational problems, such as the difference between polynomial and exponential time.
Amongst the historical examples of combinatorial problems, the ``Bridges of K\"oningsberg'' problem is a famous one solved by Euler, where the premises of computational complexity can be already sketched. This problem deals with finding a path through all parts of the city of K\"oningsberg using each bridge only once. It was solved quickly by Euler who, by using the dual graph, showed that a solution exists only if the degree of each node is even (except from the starting and ending node). This perspective is a profound change of how to view the problem since by Euler's argument, the verification of the existence of such a path can be done very quickly (in a polynomial time since it is enough to check the degree of all edges) as opposed to an exhaustive search through all possible paths.
	\begin{figure}[h]
	\centering
		\includegraphics[width=\figurewidth]{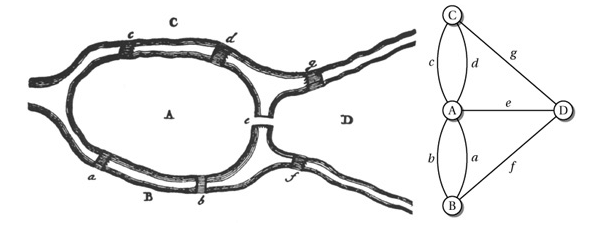}
		\caption{from \cite{Moore11}}
	\end{figure}

Hamilton came later with a problem which looks similar. It is about the possibility of finding a path in a graph that visits each vertex exactly once. Although it looks similar, the fundamental difference with the previous problem is the impossibility (as far as we know) to verify the existence of such a path by a simple (and quick) argument. Therefore finding a Hamiltonian path seems to be possible only by an exhaustive search (with an exponential number of possibilities). On the other hand, checking if a given path is a Hamiltonian cycle or not is an easy problem. One simply goes through the path, node by node, to check whether a path is a solution of the problem or not.

To go from these first simple examples to the theory of computational complexity, we first need a few definitions.

\begin{definition}[Problem] A problem is given by defining an input (or instance), for example in the case of the Hamilton cycle, the input is: ``a graph G'', and a question on that input, ``does there exist a Hamiltonian path on this graph? (yes or no)''
\end{definition}

\begin{definition}[The class P]
P is the class of problems that can be ``solved'' in polynomial time ${\rm poly}(x)=x^c$ for some constant $c$ and where $x$ is the size of the instance. In the example above, the Euler Path belongs to P as, given a graph G, we can answer the question: ``does there exist an Eulerian path on G ?'' in polynomial time.
\end{definition}

\begin{definition}[The class NP]
  NP is a class of decision problems, which have a yes/no answer. If the answer is ``yes'', there is a proof that the answer is ``yes'' that can be checked in polynomial time. 
\end{definition}

Let's look at the Hamiltonian case.  Consider the question: ``Is there a Hamiltonian path?''  We do not know any polynomial algorithm answer, but if someone provides us such a path, it is a proof that the answer is yes, and it can be checked in polynomial time.

Note that the definition of NP problems is not symmetrical. If we claim ``Prove that there is not a Hamiltonian path!'', it does not seem easy at all to prove that such a path does not exist: as far as we know, in general all possible paths have to be examined.

\begin{definition}[The class NP-complete]
 A problem is called NP-complete if it ``somehow'' encompasses any problem in NP.
\end{definition}

To be more precise we need to introduce the notion of ``reduction'' of a problem into another. A problem (A) can be reduced to (B) if, given an instance $x$ of A, there exists a function $f$ so that $f(x)$ is an instance of (B) and $f$ is in P. In addition, $f(x)$ is a ``yes''-instance of (B) if and only if $x$ is a ``yes''-instance of (A). Such a reduction is written as $A \leq B$. In practice, it means that if I have an algorithm to solve (B), I can use it in some way to solve (A). It also implies that if (B) is in P, (A) is in P since $f$ is a polynomial function. And as can be understood by the definition, if any problem of the NP-complete class is in P, then P=NP.

Here's an example of a reduction between two problems in P.

\begin{example}[Reduction--- Perfect bipartite matching $\leq$ max-flow] ~\newline The problem of telling whether a bipartite graph has a perfect matching can be mapped on the max-flow problem by adding the nodes \emph{s} and \emph{t} to the bipartite graph as illustrated below. The node \emph{s} is linked to the left part of the bipartite graph by directed edges, at the center, the edges are replaced by directed edges going from left to right and the right part is linked to the node \emph{t} by directed edges. Each edge has a weight of one. The intuition is that, if there is a max-flow of value \emph{m}, then there is a matching consisting of \emph{m} edges. By the direction of the arrow and the weight of the edges in the max flow problem, the principle of the proof can be understood.

\begin{center}
\centering
  \includegraphics[width=\figurewidth]{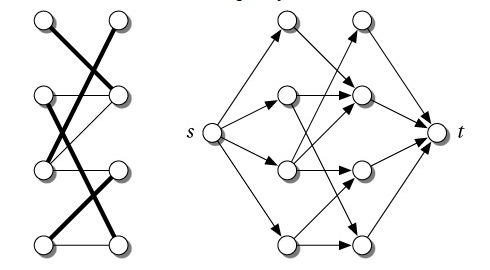}
  \centering
\end{center}

It can also be demonstrated that max-flow can be reduced to the weighted min-cut problem and vice versa.
\end{example}

The reduction of one problem into another implies also that, if A $\leq$ B then, if A is not in P then B is not in P (which justified the symbol ``less or equal''). Now we can define more clearly the class of NP-complete problems:

\begin{definition}[NP-completeness]
B is NP-complete if
\begin{itemize}
\item B $\in$ NP
\item A $\leq$ B, $\forall$ A in NP
\end{itemize}
\end{definition}

This definition implies that, if B is in P then P$=$NP, so it is enough to find a polynomial solution to an NP-complete problem to demonstrate the equality. It can be shown that the problem of finding a Hamiltonian path is NP-complete. We introduce now some examples of NP-complete problems.

\begin{example}[NP-complete problems]
\hspace{10cm}

\begin{itemize}
 \item \underline{program SAT:} A program $\Pi$ and a time $t$ given in unary. Does there exist an instance $x$ such that $\Pi(x)=$``yes'' (in \emph{t} steps or less)?
 \vspace{0.1cm}

\textup{This problem is NP-complete because it reproduces the exact structure of any NP problem: a general ``yes/no'' question that should be checkable in a time at most of size $t$. Remark: $t$ is given in unary (i.e. a string of $t$ 1s), instead of in binary. Otherwise $t$ would be exponential as a function of the size of the input, measured in bits.}

\item \underline{circuit SAT:} Boolean circuits are a set of source nodes (bits) connected to logical gates (AND,OR,NOT) on a directed acyclic graph. The output consists of a set of sink nodes. A circuit is called satisfiable if there exists an assignment of the input nodes such that the output is true. The problem can be phrased as the following:

\indent Input: A circuit C \\
\indent Question: Does there exist an x so that $C(x)$  is true?
\vspace{0.1cm}

\textup{This problem is NP-complete because the program SAT can be reduced to a circuit-SAT instance. This can be understood as any algorithm can be stated in term of a circuit-SAT form (it is enough to think that a computer is only an ensemble of nodes and logical gates, and that for-loops can be unfolded to produce more layers of these gates). Therefore, any program can be mapped to a circuit SAT instance and it is an NP-complete problem.}

\item \underline{3-SAT:} the 3-SAT problem is a constraint satisfaction problem (CSP) which is defined by an ensemble of variable nodes (true/false). Then a 3-SAT instance is a formula, which consists of a set of clauses all connected by an OR operator. A clause contains three variables linked by an AND operator and each variable can be negated or not. We can prove the following: circuit-SAT $\leq$ 3-SAT.
\vspace{0.1cm}

\textup{This reduction can be proven by first demonstrating that any circuit-SAT can be mapped into a SAT formula, and then showing that any SAT formula can be mapped in a 3-SAT formula.}
\smallskip

\textup{\textbf{Circuit-SAT to SAT formulas:} To map a circuit to a formula, we add variables for the internal values on the wires, and then transform each logical gate into a set of SAT formulas. Then the output of the gate is used as a new variable that is propagated throughout the rest of the circuit.}
\begin{itemize}
\item 1) \textbf{AND}: $y=x_1 \wedge x_2 \Leftrightarrow (x_1 \vee \bar{y})\wedge(x_2 \vee \bar{y})\wedge(\bar{x_1} \vee \bar{x_2} \vee y)$.
\item 2) \textbf{OR}: $y=x_1 \vee x_2 \Leftrightarrow (\bar{x_1} \vee y)\wedge(\bar{x_2} \vee y)\wedge(x_1 \vee x_2 \vee \bar{y})$.
\item 3) \textbf{NOT}: $y=\bar{x} \Leftrightarrow (x\vee y)\wedge(\bar{x}\vee\bar{y})$ 
\end{itemize}

\textup{Any k-SAT clause can be written as a 3-SAT formula. We refer to \cite{Moore11} for the case $k>3$ and show the cases $k=1,2$ here so that the reduction from a circuit-SAT to 3-SAT is complete.}

\begin{itemize}
\item \textup{A single variable} $x \Leftrightarrow (x \vee z_1 \vee z_2)\wedge(x\vee z_1 \vee \bar{z_2})\wedge(x\vee \bar{z_1}\vee z_2)\wedge(x\vee \bar{z_1} \vee \bar{z_2})$. \textup{In the 3-SAT formula above, whatever the values of $z_1$ and $z_2$ are, the expression is satisfiable if $x$ is true and unsatisfiable otherwise.}
\item \textup{Two variables} $(x\vee y) \Leftrightarrow (x\vee y \vee z) \wedge (x \vee y \vee \bar{z})$. \textup{Same as above, whatever the value $z$ takes, the expression is satisfiable only if $x$ OR $y$ is true.}
\end{itemize}

\textup{With these two pieces, it is now easy to map any instance of a circuit SAT toward a 3-SAT problem and therefore Circuit SAT $\leq$ 3-SAT.}

\item \underline{NAE-SAT} (NonAllEqual-SAT): This problem is very similar to a SAT problem with the small difference that all the variables in one clause cannot be all true (or false) at the same time. We can observe that any NAE-SAT solution is symmetric (unlike the k-SAT case). For this problem we have the reduction 3-SAT $\leq$ NAE-3-SAT. To prove this reduction, it is easier to prove first that 3-SAT $\leq$ NAE-4-SAT and then to show that NAE-3-SAT $\leq$ NAE-4-SAT. The first point can be proven by adding a variable to each clause of a 3-SAT instance. So any formula $\phi=(x_1 \vee x_2 \vee x_3) \wedge \ldots$ becomes $\phi'=(x_1,x_2,x_3,s) \wedge \ldots$, with the same variable $s$ added to every clause. If $\phi$ is satisfiable, we can satisfy $\phi'$ by setting $s$ to false. Now take a SAT-instance of $\phi'$. Because the problem is symmetric, the symmetric version of that instance will exist, and amongst those two one where $s$ is false. Therefore the rest of the variables satisfy the 3-SAT formula as well. Then it remains to convert any NAE-4-SAT formula into a NAE-3-SAT formula. This can be done easily by using another variable $z$ and seeing that $(x_i,x_j,x_k,x_l)=(x_i,x_j,z) \wedge (\bar{z},x_k,x_l)$.

\end{itemize}
\end{example}
\smallskip

\section{Hardness: P, NP or EXP?}
We already discussed the hardness of a problem in the previous lecture. Now we want to discuss counting problems. We already know  that problems in NP ask whether an object with a certain property exists. We now define $\#$P, pronounced ``sharp P'', as the class of problems that ask how many such objects exist. As in NP, we require that this property can be checked in polynomial time.
\vspace{0.5cm}

\begin{minipage}{8cm}
\begin{definition}
\textbf{$\#P$} is the class of functions $A (x)$ of the form
$A (x)= \#\{w \mid B(x, w)\}$,
where $B (x, w)$ is a property that can be checked in polynomial time, and where $|w | = poly(|x |)$ for all $w$ such that $B(x, w)$ holds.
\end{definition}
\end{minipage}
\begin{minipage}{0.5cm}
${}$
\end{minipage}
\begin{minipage}{6cm}
\centering
\includegraphics[width=3.5cm]{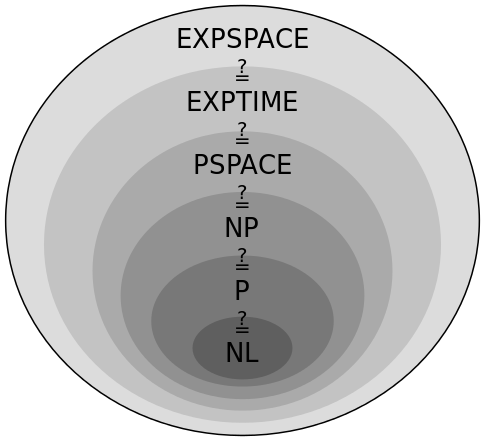}
\begin{center}
\small{from \cite{Moore11}}
\end{center}
\end{minipage}

\subsection{Examples}
\begin{example}[Perfect Matching]
\label{ex:perfectMatching}
 The problem of deciding if there exists a perfect matching in a given graph $G$, is, as we already mentioned, in P. On the other hand the problem of counting these perfect matchings is in $\#$P. Suppose we have a bipartite graph $G$ with $2n$ vertices. We can represent it as a $(n\times n)$ matrix $A$, where $A_{ij} = 1$ if the $i$th vertex on the left is connected to the $j$th vertex on the right, and $A_{ij} = 0$ otherwise. \\

\begin{minipage}{7.5cm}
\includegraphics[width=6cm]{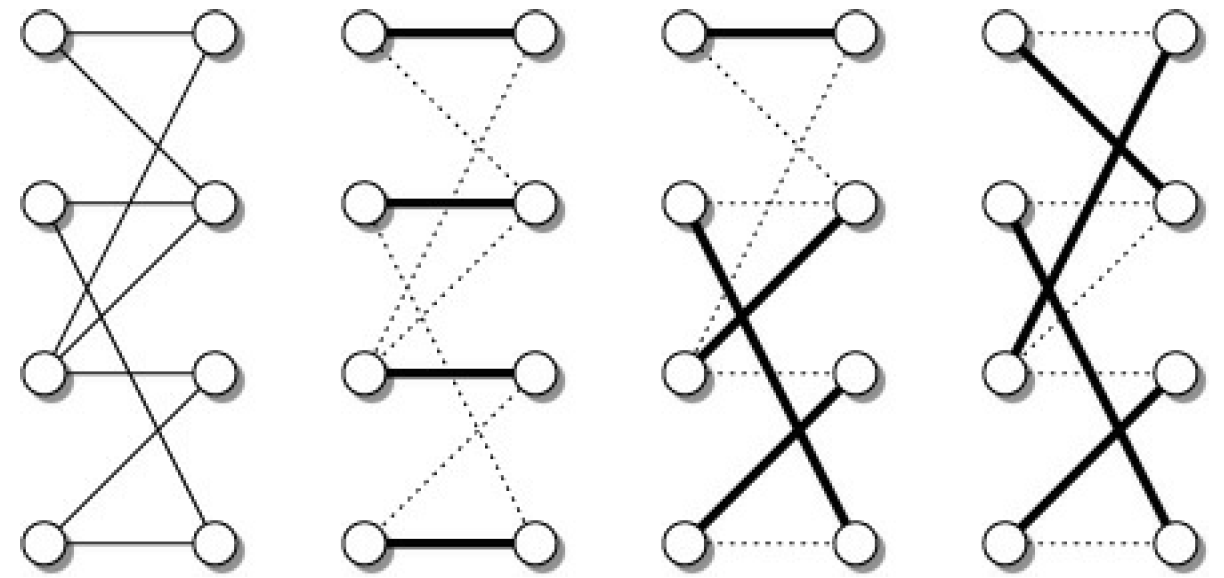}
\begin{center}
 from \cite{Moore11}
\end{center}

\end{minipage}
\begin{minipage}{6.5cm}
As an example we can write down the matrix $A$ for the first graph on the left:

\begin{center}
$A=
\begin{pmatrix}
	1 &1&0&0 \\
0	&1&0&1 \\
1	&1&1&0 \\
	0&0&1&1
\end{pmatrix}$
\end{center}

\end{minipage}

\vspace{0.2cm}
But how can we express the number of perfect matching in terms of $A$? Each perfect matching is a permutation that maps the vertices on the left to their partners on the right. Each permutation $\sigma$ corresponds to a matching if and only if for each $i$ there is an edge from $i$ to $\sigma(i)$, i.e., if $A_{i\sigma(i)} = 1$. Therefore, the number of matchings is given by the following quantity, which is called the \textbf{permanent of A}:
\begin{center}
$\perm(A)=\sum_{\sigma\in S_n}{\prod_{i=1}^{n}{A_{i\sigma(i)}}}$
\end{center}
Note that this is just like the determinant, except that it doesn't have the parity $(-1)^\sigma$.  Ironically, this makes it harder to compute.
\end{example}
\begin{example}[Graph 3-colorability]
We want to give a hint on how it is possible to reduce graph 3-colorability to planar graph 3-colorability. In other words how we can transform an arbitrary graph $G$ to a planar graph $G'$, such that $G'$ is 3-colorable if and only if $G$ is. We can easily see that the reverse reduction is trivial. Thus planar graph 3-colorability is just as hard as graph 3-colorability in general, and the two problems have the same complexity.\smallskip

\begin{minipage}{7.5cm}
\includegraphics[width=6cm]{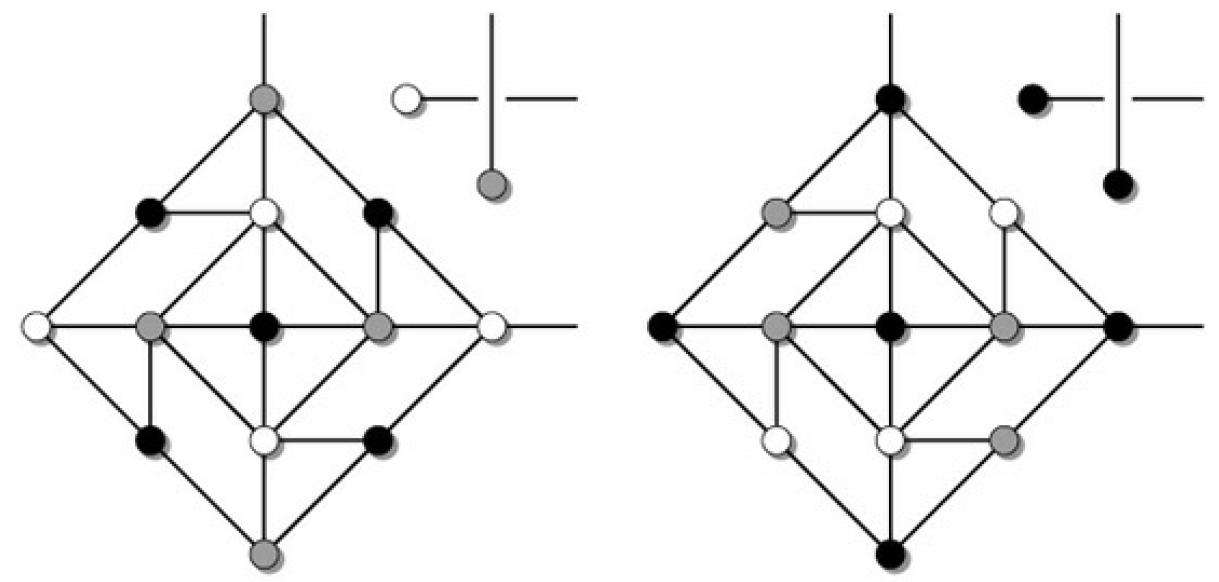}
\begin{center}
 from \cite{Moore11}
\end{center}
\end{minipage}
\begin{minipage}{6.5cm}
This crossover gadget allows us to convert an arbitrary graph into a planar one while preserving the property of 3-colorability. The gadget can be colored in two different ways up to symmetry, depending on whether the transmitted colors are different (left) or the same (right).
\end{minipage}
\end{example}

\begin{example}[2-dimensional Ising model]
We want to think about the Ising model as a 2-dimensional lattice with $n$ spins which can take the values $\pm1$. Now we can consider the dual lattice, whose edges cross the  edges of our Ising model lattice.\\

\begin{minipage}{7.5cm}
 An edge in the dual lattice is colored in red if the two spins that are connected through its crossing edge have different values. It is clear that one vertex of the dual lattice is always adjacent to an even number of red edges. Now we can think about a ground state of the Ising model as a \textsl{minimum weight perfect matching} in a decorated version of the dual graph (see~\cite{Moore11} for the gadgets and details). We know that every spin assignment corresponds to a single coloring of the dual lattice.   
\end{minipage}
\begin{minipage}{6.5cm} 

\begin{flushright}
\includegraphics[width=5cm]{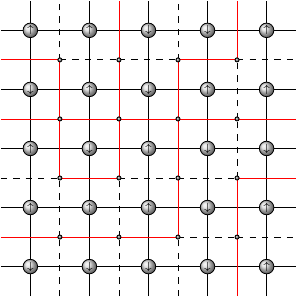}
\end{flushright}


\end{minipage}
\end{example}

\begin{example}[Domino tiling]\smallskip
${}$

\begin{minipage}{7.5cm}

Our next task is to reduce domino tiling (where we want to cover a $n\times n$-chess board $C$ with dominos) to finding a perfect matching in a bipartite graph. We therefore define a graph $G$ where each vertex corresponds to a square in $C$ and is connected to the four vertices corresponding to the neighboring squares. A domino tiling of $C$ is just a perfect matching of $G$. So domino tiling is in P.

\end{minipage}
\begin{minipage}{6.5cm}
\centering
 \includegraphics[width=3cm]{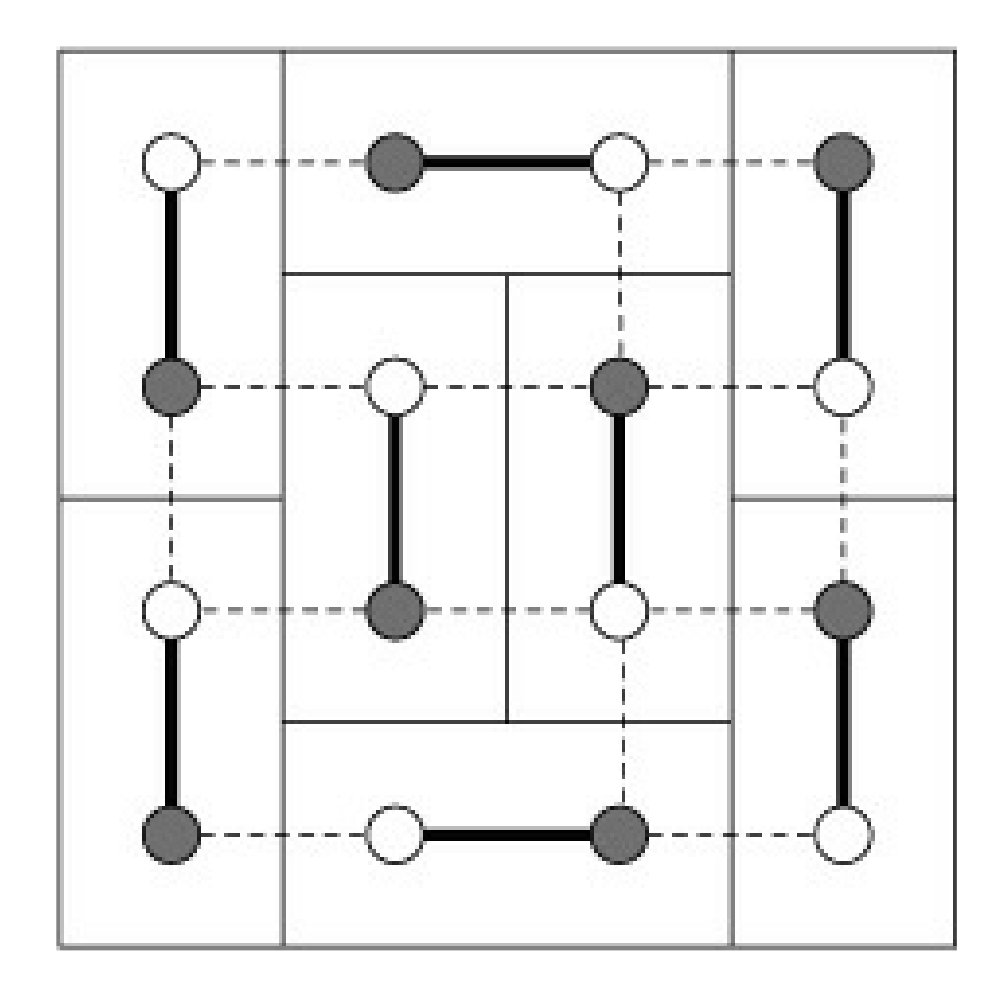}

\begin{center}
 from \cite{Moore11}
\end{center}
\end{minipage}
\smallskip

\begin{minipage}{7.5cm}
\centering
\includegraphics[width=6cm]{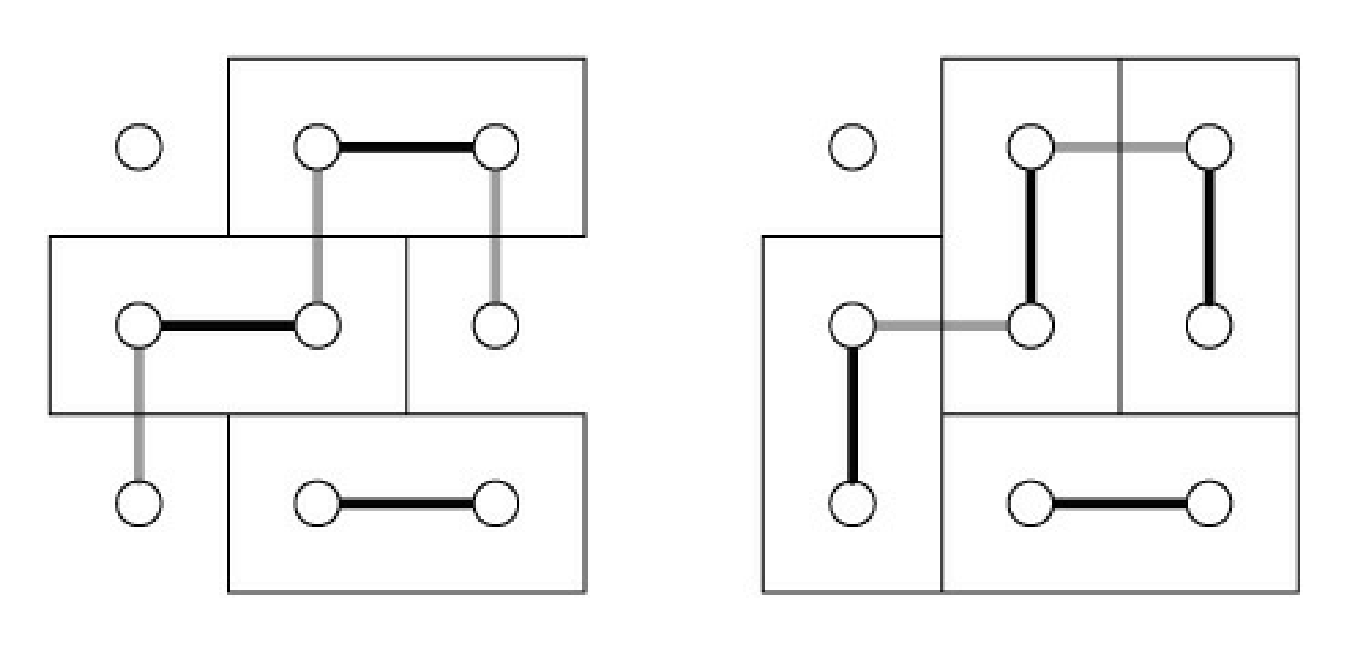}
\begin{center}
 from \cite{Moore11}
\end{center}
\end{minipage}
\begin{minipage}{6.5cm}
We can improve a partial domino tiling, or equivalently a partial matching, by flipping the edges along an alternating path. Now if we want to compute the number of domino tilings for the matrix $C$ we have to compute $\perm(C)$ as above.
\end{minipage}
\end{example}

\textbf{Turning Permanents into Determinants} Again we have a planar, bipartite graph $G$ with $n$ vertices of each type. Now we color the two types of vertices in black and white (see above). As in example \ref{ex:perfectMatching} we define a $n\times n$-matrix $A$. 
We know already that each perfect matching corresponds to a permutation $\sigma\in S_n$, which maps each black vertex to its white partner. We saw in example \ref{ex:perfectMatching} that $\perm(A)$ counts all of these permutations. But the determinant of $A$ counts them weighted by their parities, $\det(A) =\sum_{\text{matchings }\sigma}(-1)^{\sigma}$. The idea is to compensate the parity weights of the determinant in order to obtain the correct count of perfect matching. To do this, we place weights $w_{ij}=\pm 1$ on the edges of $G$. It defines a quantity $\tilde{A}_{ij}=w_{ij}$. Now, each matching $\sigma$ of $\tilde{A}$ has a weight 
\[w(\sigma)=\prod_i w_{i\sigma(i)}\]
and the determinant of $\tilde{A}$ is given by
\[\det(\tilde{A})=\sum_{\text{matchings }\sigma}(-1)^{\sigma}w(\sigma)\]
Now we would like to write the permanent of $A$ as the determinant of $\tilde{A}$. To do this, we should choose the weights $w_{ij}$ such that $(-1)^\sigma w(\sigma)$ has the same sign for all $\sigma$. It means that the matching $\sigma$ should change the weight $w(\sigma)$ by a factor $-1$ when its parity changes. Then we would have that $|\det(\tilde{A})|=\perm(A)=\#$ of perfect matchings. The whole trick is to find a proper set of weights. For instance, in the particular case of the chess board (see the figure above), one can decide to put a weight $i$ on all vertical edges. Then, by changing two horizontal dimers to two vertical ones, the weight changes by $i^2 = -1$, and so does the parity. It should be emphasized that a set of such weights can always be found for planar graphs. In addition, an analogous approach allows us to solve the 2-d Ising model (using a trick to count perfect matchings on planar graph that are not bipartite). For further details we refer to \cite{Moore11}. We finally notice that some problems (like Graph 3-Coloring) are just as hard when limited to the planar case, while others (like 4-Coloring or counting matchings) get much easier.

We new turn to random graphs, and give some of their basic properties. These properties will be useful when discussing some other problems in more details.

\section{Random Graphs}

\subsection{Erd\H{o}s-R\'{e}nyi Random Graphs}
First we want to give the definition of a random graph.

\begin{definition}
A graph $G(n,p)$ with $n$ vertices and edge probability $p$ is one model of a \textbf{random graph}.  Edge probability $p$ means that between any two vertices \emph{v} and \emph{w} there exists an edge with probability $p$. For every pair of vertices this event is independent from each other pair. An \textbf{Erd\H{o}s-R\'{e}nyi graph} $G(n,m)$ has $n$ vertices and is uniformly chosen from all graphs with exactly $m$ edges. 
\end{definition}

Our first task is to compute the expected degree of one vertex and the expected number of edges in the $G(n,p)$ model. We want to investigate the sparse case $pn=c$ where $c$ is a constant independent of $n$.  We use $\mathbb{E}$ to denote the expectation.

\begin{align}
&\mathbb{E}[m]=p\binom{n}{2} \approx \frac{pn^2}{2}=\frac{cn}{2}\\
&\mathbb{E}[deg]=p(n-1)\approx pn=c
\end{align}

Now we can compute some other interesting quantities in our model. We compute the expected number of triangles and bicycles in $G(n,p)$,
\begin{align}
&\mathbb{E}[\#\Delta]=\binom{n}{3}p^3\approx\frac{c^3}{6}\\
&\mathbb{E}[\#\text{ bicycles}]=const\times\binom{n}{k}p^{k+1}\approx n^k\frac{c^{k+1}}{n^{k+1}}\stackrel{n\rightarrow\infty}\longrightarrow 0
\end{align} 
Bicycles are subgraphs containing $k$ vertices and $k+1$ edges.  Thus they contain two loops, connected by a path or by an edge belonging to both loops. The constant depends on $k$ but not on $n$.
For any fixed $k$ there are probably  no bicycles as $n \to \infty$, so most vertices have a treelike neighborhood. 

Our next task is to compute the probability that a vertex $v$ has degree $k$.
\begin{align}
\P[\mathrm{deg}(v)=k]=\binom{n-1}{k}p^k(1-p)^{n-1-k}\approx\frac{n^k}{k!}\frac{c^k}{n^k} \left( 1-\frac{c}{n} \right)^n=\frac{\e^{-c}c^k}{k!}
\end{align}
This is called the \textbf{degree distribution}. As $n$ goes to infinity, it becomes a Poisson distribution with mean $c$.

\subsection{The Giant Component}

In this section we want to describe the most basic phase transition occurring in $G(n, p)$ for $pn=c$. For very small $c$, $G$ consists of small components isolated from each other, and almost all of them are trees. For $c$ bigger and bigger we add more edges and these trees grow and connect with each other. Suddenly, at $c = 1$, many of them come together to form a giant component. We want to observe this phase transition later on in this chapter. First we study the expected component size of a vertex $v$. We start at this vertex and go further to its neighbors and their neighbors and so on. We can think of those vertices as $v$'s children and the descendants with distance $k$ from our starting vertex $v$ are the so called \textbf{$k^{\rm th}$ generation} of $v$. By doing that we can develop the whole component of $v$. For large $n$ we can assume that every child of $v$ has again Poisson($c$) children.  That is, we can approximate the process of exploring $v$'s component as a branching process.  We only have to count the descendants of $v$ to get the component size.
\begin{align}
\mathbb{E}[\text{Component size of }v]=1+c+c^2+c^3+ \cdots =\frac{1}{1-c} \text{ for } c<1
\end{align}
For $c<1$ this sum converges and the expected component size is finite. For $c>1$, on the other hand, the sum diverges and $v$ has, in expectation, an infinite number of descendants; in fact, the number of descendants is infinite with positive probability. At this point the branching process is no longer an accurate model to compute the component size of $v$. For further computations we call $\gamma$ the fraction of all vertices that lies in the giant component of $G(n,p)$. When $c > 1$, with high probability there is a unique giant component.

We will see two different methods to estimate the value of $\gamma$.
\subsubsection{Method \#1}
The probability that a vertex $v$ is not part of the giant component is the sum over all $k$ of the probability that $v$ has degree $k$ and none of its children are part of the giant component.

\begin{align}
1-\gamma=\sum_k \P[\mathrm{deg}(v)=k](1-\gamma)^k
= \sum{k}\frac{\e^{-c}c^k}{k!}(1-\gamma)^k
= \e^{-c} \e^{c(1-\gamma)}
= \e^{-c\gamma}
\end{align}

\begin{minipage}{5cm}
So we get a transcendental equation for the size of the giant component:
\begin{align}
\gamma=1-\e^{-c\gamma}
\end{align}
\end{minipage}
\begin{minipage}{8cm}

\centering
\includegraphics[width=10cm]{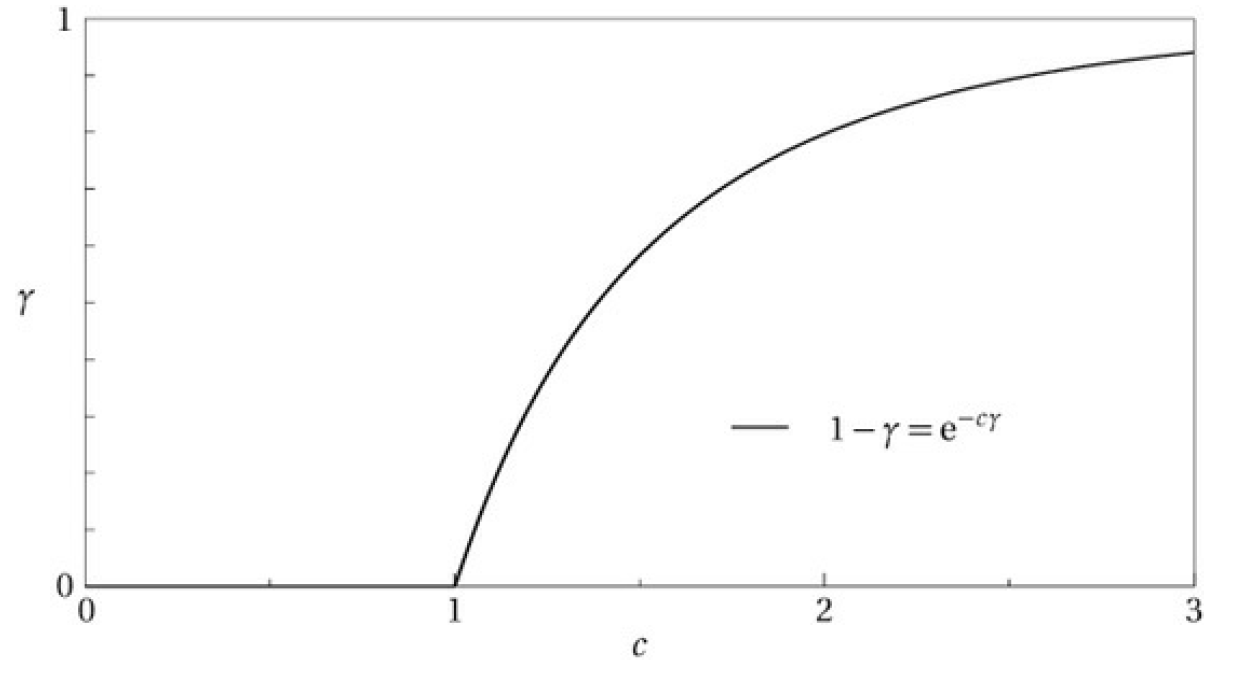}
\begin{center}
 from \cite{Moore11}
\end{center}
\end{minipage}

\subsubsection{Method \#2}
We compute $\gamma$ again but now with a system of differential equations. We start again with a vertex $v$ and explore its connected component, now by using an algorithm. At each point in time, a vertex is labeled \textbf{Reached}, \textbf{Boundary}, or \textbf{Untouched}. 
\begin{itemize}
	\item ``Reached'' means that a vertex lies in $v$'s component and its neighborhood has been explored 
	\item ``Boundary'' means that a vertex lies in $v$'s component but its neighbors have not been explored yet
	\item ``Untouched'' means that it is not yet known if the vertex is in $v$'s component.
	\end{itemize}
	
\begin{algorithm}
\caption{Cluster expansion}
\begin{algorithmic}[1]
\REQUIRE a vertex $v$
\ENSURE a connected-graph

\STATE label $v$ Boundary
\STATE label all other vertices Untouched
\WHILE{there are Boundary vertices}
\STATE choose a Boundary vertex
\STATE label each of $v$'s Untouched neighbors Boundary 
\STATE label $v$ Reached
\ENDWHILE
\end{algorithmic}
\end{algorithm}
This algorithm explores $G$ one vertex at a time, until there are no Boundary vertices left and $v$'s component has been labeled Reached.
Let $R$, $B$, and $U$ denote the number of vertices of each type at a given point in time. At each step of the algorithm, $R$ increases by 1, and the expected change in $B$ and $U$ is
\begin{align}
&\Delta R=1\\
&\mathbb{E}[\Delta B]=-1+pU=\frac{c}{n}U-1\\
&\mathbb{E}[\Delta U]=-pU=-\frac{c}{n}U
\end{align}
We have an expected change from $U$ to $B$ of $pU$, since each Untouched vertex is connected to $v$ with probability $p$. By changing the chosen vertex $v$ from Boundary to Reached we get the term $-1$.  When $n$ is large we can rescale these stochastic difference equations, so they become a system of differential equations:
\begin{align}
&\frac{\dr}{\dt}=1\\
&\frac{\db}{\dt}=cu-1\\
&\frac{\du}{\dt}=-cu
\end{align}
We solve these differential equations with the initial conditions $b(0) = 0$ and $u(0)= 1$:
\begin{align}
u(t)=\e^{-ct} \text{ and } b(t)=1-t-\e^{-ct}
\end{align}
The fraction $\gamma$ of vertices in the giant component is the value of $t$ at which no Boundary vertices remain and the algorithm stops. This gives the same equation as before,  
\begin{align}
b(\gamma)=1-\gamma-\e^{-c\gamma}=0
\end{align}
Remark: The differential equation approach to the size of the giant component in random graphs is similar to the dynamics for the SIR model, where $p$ is the transmission rate and $\gamma$ is the fraction of the population that eventually becomes infected.
\medskip

Another interesting quantity we can consider is the expected number of components/trees of $G$ with $k$ vertices.  We approximate this with the number of trees:
\begin{align}
\mathbb{E}[\text{Components/trees of size }k]=\binom{n}{k}p^{k-1}(1-p)k^{k-2}(1-p)^{kn-k}
\end{align} 
Here $k^{k-2}$ comes from Cayley's formula for the number of labeled trees with $k$ vertices.
For $p=c/n$ and $c=1$, the expression above becomes a power law by applying Stirling's formula:
\begin{align}
\mathbb{E}[\text{Components/trees of size }k]\approx\frac{1}{\sqrt{2\pi}}\frac{n}{k^{5/2}}\left(1+O(k^2/n)\right)
\end{align}
 
\subsection{Giant component and configuration model}

One problem of random graphs is that the degree distribution converges toward the Poisson distribution which is unrealistic in many concrete examples (social networks, biological networks, etc.). The configuration model can be used to remedy this problem. This model deals with a sequence of nodes and degrees. The nodes are chosen randomly before being connected by an edge. During this process, each node is chosen with a probability proportional to the number of unmatched edges it has.  Equivalently, each vertex with degree $d$ has $d$ ``stubs'' or half-edges.  The total number of stubs is $2m$ where $m$ is the number of edges, and we choose a uniformly random matching of these stubs with each other.  
\smallskip 

This procedure may create some self-loops or multiple edges, so strictly speaking this model produces random multigraphs.  However, for reasonable degree distributions (with bounded mean and variance) the resulting graph is simple with constant probability.

\begin{algorithm}
\caption{Configurational model}
\begin{algorithmic}[1]
\REQUIRE degree sequence
\ENSURE a graph

\WHILE{there remain unmatched edges} 
\STATE choose two unmatched edges uniformly and randomly 
\STATE put an edge between them 
\ENDWHILE
\end{algorithmic}
\end{algorithm}

The next question is to determine where, in this new setting, the giant component appears.
As long as the graph is sparse (i.e. the average degree is constant) the appearance of the giant component can be analyzed by a branching process as before. 
When we follow a link to a vertex of degree $k$, there are $k-1$ ``children'' consisting of new edges that come out of it.  
Thus the average branching ratio is
\begin{equation}
	\lambda = \sum_k (k-1) \frac{k a_k}{\sum j a_j} 
	= \frac{\mathbb{E}[k(k-1)]}{\mathbb{E}[k]}
\end{equation}
where $a_k$ is the fraction of nodes in the degree sequence that have degree $k$.  
A giant component appears when $\lambda > 1$, or equivalently
\begin{equation}
    \mathbb{E}[k^2] > 2 \mathbb{E}[k]
\end{equation}
Thus both the first and second moments of the degree distribution matter.

\subsection{The $k$-core}
\begin{definition}
The \textbf{$k$-core} of a graph $G$ is the largest sub-graph where each vertex has minimal degree $k$. Equivalently, it is the graph remaining after removing vertices with $d<k$ neighbors (iteratively).
\end{definition}
This object is related to the $k$-colorable property of a graph: if there is no $k$-core, the graph is $k$-colorable \cite{pittel1996sudden} (note that the converse is not necessarily true). We will again use a branching process to characterize the $k$-core: at which $\alpha$ it appears and what fraction of the system is in it at that point. Unlike the giant component, the $k$-core appears suddenly for $k \ge 3$: that is, it includes a constant fraction of vertices when it first appears.

\paragraph{Branching process} We consider only $k>2$ since for any $c > 0$, $G(n,p=c/n)$ typically contains a $2$-core as a loop of size $\mathcal{O}(\log n)$ or even a triangle is present with constant probability.  
To describe the problem with a branching process, we need to consider first a root node $v$, again treating its neighbors as its children, their neighbors as its grandchildren, and so on.  Let us (recursively) say that a node in this tree is \emph{well-connected} if it has at least $k-1$ well-connected children.  Such a node will survive the process that deletes nodes with degree less than $k$, if we think of this process as moving up the tree from the leaves.  

If the fraction of well-connected children is $q$, the number of well-connected children a given node has is Poisson distributed with mean $cq$. The probability that the number of well-connected children is less than $k$ is then given by
\begin{equation}
  Q_k = \sum_{j=0}^{k-1} \frac{\e^{-cq} (cq)^j}{j!}.
\end{equation}
This gives us the fixed-point equation
%
%
\begin{equation}
\label{eq:core}
 q=1-Q_{k-1} \, .
\end{equation}
With this equation, we can find the value of $c$ at which the $k$-core appears, i.e., the smallest $c$ such that this equation has a positive root. For $k=3$, for instance, we have $c^{\rm{CORE}}_3=3.351$.  

The probability that a given node is in the core is a little tricky.  Here we treat the node as the root. In order to survive the deletion process, it has to have at least $k$ well-connected children.  Thus the fraction of nodes in the core is 
\begin{equation}
\gamma_k = 1-Q_k \, . 
\end{equation}
where $q$ is the root of~\eqref{eq:core}.  For $k=3$ this gives $\gamma_3=0.268$.

\section{Random $k$-SAT}

In analogy with random graphs $G(n,m)$ that have $n$ nodes and $m$ edges, we can define random $k$-SAT formulas $F_k(n,m)$ with $n$ variables and $m$ clauses.  We choose each clause uniformly and independently from the $2^k {n \choose k}$ possible clauses.  That is, we choose a $k$-tuple of variables, and then for each one independently negate it with probability $1/2$.


We focus on the sparse case where the number of clauses scales as $m=\alpha n$ where $\alpha$ is a constant. In the following analysis we will not care about rare events such as two variables appearing in the same clause, or the same clause appearing twice in the formula.  

\underline{Exercise}: show that in the sparse case, the probability of either of these happening tends to zero as $n \to \infty$.

\vspace{0.2cm}
\begin{minipage}{5cm}
\textbf{Phase Diagram:} It is now commonly accepted that the random $k$-SAT (here $k=3$) problem undergoes a phase transition at a critical value $\alpha_k$. This phase transition is characterized by the probability of having a satisfying assignment converging to 1 below $\alpha_k$ and to 0 above. 
\end{minipage}
\begin{minipage}{0.6cm}
${}$
\end{minipage}
\begin{minipage}{10cm}
\centering
\includegraphics[width=8cm]{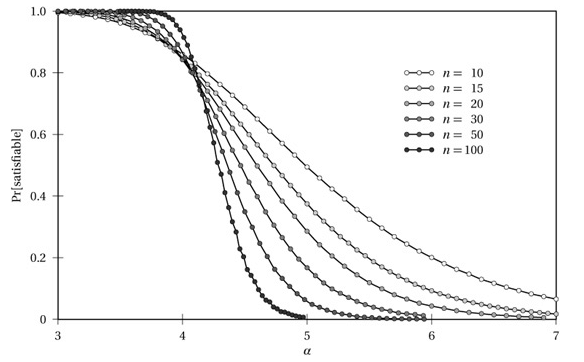}
\begin{center}
from \cite{Moore11}
\end{center}
\end{minipage}

\vspace{0.5cm}

Mathematically speaking, this picture is still a conjecture which is known as:\smallskip
\begin{conjecture}[Threshold conjecture]
\begin{center} 
\indent $\forall k \geq 3$, $\exists \alpha_k$ such that $\forall \epsilon>0$, \\
\medskip
\indent $\lim \limits_{n \to \infty} \P[F_k(n,\alpha m) \text{ is satisfiable}] = \left\{ 
\begin{array}{ll}
  1 & \mbox{if } \alpha < (1-\epsilon) \alpha_k \\
  0 & \mbox{if } \alpha > (1+\epsilon) \alpha_k 
\end{array} \right.$
\end{center}
\end{conjecture}
The closest rigorous result for this conjecture came from Friedgut, who showed that there is a function $\alpha_k(n)$ for which it is true. But we don't know if $\alpha_k(n)$ converges to a constant as $n \to \infty$, or whether it continues to fluctuate in some way.  Physically, this would be like the freezing point of water depending on the number of water molecules, which seems very unlikely, but we still lack a rigorous proof.  
For the rest of this section, we will assume that the threshold conjecture is true.

\subsection{Easy Upper Bound} 

It is easy in a first approach to derive a (not so good) upper bound for the position of the $k$-SAT threshold.  We will use the ``first moment'' method: for any random variable that takes values $0, 1, 2, \ldots$, such as the number of some object, we have (exercise)
\[
\P[x>0] \leq \mathbb{E}[x] \, . 
\]
Let's define $x$ as the number of satisfying assignments for a given formula.  
It is easy to compute the average of $x$ as the clauses are independent. Let write $x$ as the sum over all truth assignments or ``configurations'', using an indicator function to count only the satisfying ones:
\begin{eqnarray}
	x & = & \sum_{\sigma \in \{0,1\}^n} \mathbbm{1}_\sigma \\
	\mathbb{E}[x] & = & \sum_\sigma \mathbb{E} [ \mathbbm{1}_\sigma ] \\
	& = & \sum_\sigma \P(\sigma \;\text{satisfies}\; \phi) = \sum_\sigma \prod_{c \in \phi} \P(\sigma \;\text{satisfies}\; c) \\
	& = & 2^n (1-2^{-k})^m \, . 
\end{eqnarray}
For $k=3$ in particular, this gives
\[
\mathbb{E}[x] = \left[ 2 \left(\frac{7}{8}\right)^\alpha \right]^n \, . 
\]
Therefore we can compute an upper bound on $\alpha_3$ by finding $\alpha$ such that $2(\frac{7}{8})^\alpha = 1$.  This gives $\alpha_3 \le 5.19$.  More generally, we have 
\[
\alpha_k < \frac{\ln 2}{\ln (1-2^{-k})} < 2^k \ln 2 \, . 
\]
An interesting question, which was open until fairly recently, is whether this bound is asymptotically tight.  How close to the truth is this simple counting argument?  For that matter, why is it not exactly correct?  For $k=3$, in the range $4.27 < \alpha < 5.19$, the expected number of satisfying assignments is exponentially large, and yet most of the time there aren't any at all.  What's going on?

\subsection{Lower Bounds from Differential Equations and the Analysis of Algorithms}

We will go through two proofs for the lower bound of the $k$-SAT threshold. The first (easy) one is based on the analysis of a very simple algorithm called ``Unit Clause propagation'' (UC). This algorithm deals with unit clauses which are clauses containing a single variable ($x$ of $\bar{x}$). How could unit clauses appear in the $k$-SAT problem ? Well, if one fixes a variable of the problem, three different options can happen concerning the clauses in which it appears. If it satisfies a clause then the clause disappears. If the clause remains unsatisfied, then the variable is removed from the clause. In that case, a 3-clause becomes a 2-clause, and a 2-clause becomes a unit clause. If the clause in which the variable appears is a unit clause and does not get satifisfied when fixing the variable, then a contradiction appears.

The principle of UC is the following. Whenever there exists a unit clause, the algorithm chooses a unit clause uniformly at random and satisfies it by permanently fixing the variable of the clause. If no more unit clauses are present, the algorithm chooses a variable uniformly at random from the unset variables and fixes it randomly to $0$ or $1$ with probability $1/2$. When all the clauses have been satisfied, the algorithm returns ``satisfiable''. If the process encounters a contradiction (unsatisfied clause with all the variables of the clause fixed), it returns ``contradiction'' and the algorithm stops.


\begin{algorithm}
\caption{Unit Clause propagation (UC)}
\begin{algorithmic}[1]

\REQUIRE a k-SAT instance
\ENSURE  ``satisfiable''  or ``contradiction''

\WHILE{(there is no contradiction) AND (there exists unsatisfied clauses)}	
\IF{there exists a unit clause}
\STATE (forced step) choose one uniformly at random and satisfy it
\ELSE
\STATE (free step) choose $x$ uniformly from all the unset variables $\rightarrow$ set $x=0,1$ 
\ENDIF
\IF{there exists an empty clause (or unsat clause)}
\STATE \textbf{return} ``contradiction''
\ENDIF
\ENDWHILE

\STATE \textbf{return} ``satisfiable''
\end{algorithmic}
\end{algorithm}
At all times, the remaining formula is uniformly random conditioned on the number $S_3, S_2, S_1$ of unsatisfied $3$-clauses, $2$-clauses, and unit clauses.  
This comes from the fact that, since variables are removed from the formula whenever they are set, we know nothing at all about how the  unset variables appear in the remaining clauses.
Moreover, both moves effectively give a random variable a random value, since the (free) move is completely random, and the forced one satisfies a random unit clause. This property makes it easier to deal with the algorithm. 

Let's imagine now that we are at a time $t$ where there remains $n-T$ variables. When dealing with 3-clauses, the probability that a chosen variable appears in it is $3/(n-T)$. When fixing this variable, there is half a chance that it will satisfy the clause and half a chance that the 3-clause became a 2-clause. The same calculation can be done for the 2-clause, a variable having $2/(n-T)$ chance to be in it.  
If we write $S_3 = s_3 n$, $S_2 = s_2 n$, and $T=tn$, and assume that the change in these rescaled variables equals its expectation, 
we can write a set of differential equations for $s_2$ and $s_3$:
\begin{eqnarray}
	\frac{\ds_3}{\dt} & = & - \frac{3 s_3}{1-t} \\
	\frac{\ds_2}{\dt} & = & - \frac{2 s_2}{1-t} + \frac{3}{2}\frac{s_3}{1-t} \, . 
\end{eqnarray}
The solution of these equations is
\begin{eqnarray}
	s_3 & = & \alpha (1-t)^3 \\
	s_2 & = & \frac{3}{2} \alpha t (1-t)^2
\end{eqnarray}
These equations don't tell us how many unit clauses there are.  Since there are $O(1)$ of these as opposed to $O(n)$ of them (indeed, there is a contradiction with high probability as soon as there are $O(\sqrt{n})$ of them) we will model them in a different way, using a branching process rather than a differential equation. 

Each time we satisfy a unit clause, we might create some new ones by shortening a $2$-clause.  Thus the branching ratio, i.e. the expected number of new unit clauses, is $\lambda=s_2/(1-t)$.  If $\lambda > 1$, this branching process explodes, leading to a contradiction.  On the other hand, it can be shown that if $\lambda < 1$ throughout the algorithm, then it succeeds in satisfying the entire formula with positive probability. Because of Friedgut's theorem, proving positive probability is enough to prove probability $1$.
Therefore we should determine the maximum value $\alpha$ can take such that $\max_t \lambda \le 1$.  We find that 
\begin{equation}
	\lambda_{\max} = \frac{3}{8}\alpha \, ,
\end{equation}
and therefore 
\[
\alpha_3  > \frac{3}{8} \, . 
\]
This analysis can be generalized for any $k$, giving
\begin{equation}
\alpha_k \gtrsim \frac{2^k}{k} \, .
\end{equation}
Note that this is a factor of $k$ below the first-moment upper bound.  Next, we will see how to close this gap.

\subsection{Lower bounds from the second moment method} 

We now use the so-called second moment method to narrow the gap between the two bounds on $\alpha_k$ previously discussed. The second moment method relies on the inequality
\begin{equation}
	\P[X>0]\geq\frac{\E[X]^2}{\E[X^2]}.
	\label{eq_ineq1}
\end{equation}
$X$ denotes the number of solutions to the formula $\phi$. 
Note that by Friedgut's theorem, we will automatically have $\P[X>0]=1$ by simply proving that the RHS is strictly positive. We will now prove the following lower bound on $\alpha_k$:
\begin{equation}
\alpha_k\geq2^{k-1}\ln2-O(1)
\end{equation}
This will reduce the width of the gap between upper and lower bound, from a factor of $k$ to a factor of $2$, independent of $k$. 
To do so, the new challenge is to compute $\E[X^2]$. Using the indicator function $\mathbbm{1_\sigma}$ introduced previously --- which is equal $1$ if $\sigma$ satisfies $\phi$ and $0$ otherwise--- we have the following:
\begin{equation}
X^2
=\Big(\sum_\sigma \mathbbm{1_\sigma}\Big)^2
=\sum_{\sigma,\tau\in\{0,1\}^n}\mathbbm{1_\sigma}\mathbbm{1_\tau} \, ,
\end{equation}
so that 
\begin{equation}
\E[X^2]=\sum_{\sigma,\tau\in\{0,1\}^n}\P[\sigma,\tau\text{ both satisfy }\phi]
\end{equation}
While using the first moment method previously, we had to compute $\E[X]$, which could be expressed in terms of $\P[\sigma \text{ satisfies } \phi]$, which was independent of $\sigma$ due to the definition of $F_k(n,m)$. On the other hand, $\P[\sigma,\tau\text{ both satisfy }\phi]$ now depends on the Hamming distance $z$ between the truth assignments $\sigma$ and $\tau$. More precisely, by independence of the clauses, we have 

\begin{align*}
\E[X^2]&=\sum_{\sigma,\tau\in\{0,1\}^n} \prod_{c\in\phi} \P[\sigma,\tau\text{ both satisfy } c]\\
&=\sum_{\sigma,\tau\in\{0,1\}^n}\Big(\P[\sigma,\tau\text{ both satisfy a random } c]\Big)^m
\end{align*}
By the inclusion-exclusion principle, we then have:
\begin{align*}
\P[\sigma,\tau\text{ both satisfy a random } c]&=1-\P[\sigma\text{ doesn't satisfy } c ]\\
&-\P[\tau\text{ doesn't satisfy } c ]+\P[\sigma,\tau\text{ both don't satisfy } c]\\
&=1-2^{-k}-2^{-k}+(z/n)^k2^{-k}\\
&=f\Big(\frac{z}{n}\Big)
\end{align*}
To compute $\P[\sigma,\tau\text{ both don't satisfy } c]$, we used the fact that
\begin{align*}
\P[\sigma,\tau\text{ both don't satisfy } c]&=\P[\tau\text{ doesn't satisfy } c ]\P[\sigma \text{ doesn't satisfy } c \lvert\ \tau \text{ doesn't satisfy } c]\\
&=2^{-k}(z/n)^k
\end{align*}
because $\sigma$ won't satisfy $c$ if and only if none of the variables that differ between $\sigma$ and $\tau$ fall in the set of variables concerned by the clause $c$. This happens with probability $(z/n)^k$. 

We therefore have
\begin{align*}
\E[X^2]
=\sum_{z=0}^n 2^n {n\choose z}f\Big(\frac{z}{n}\Big)^m
\end{align*}
where $2^n {n\choose z}$ is the number of couples of truth assignments $\sigma$ and $\tau$ such that they differ by $z$ bits. Note that when $\frac{z}{n}=\frac{1}{2}$, which is the most likely overlap between two truth assignments, then  $f\Big(\frac{z}{n}\Big)=f\Big(\frac{1}{2}\Big)=(1-2^{-k})^2$, as if $\sigma$ and $\tau$ were independent. If this term dominates the sum in $\E[X^2]$, then we have $\E[X^2]=4^n(1-2^{-k})^{2m}=\E[X]^2$, so that there exists a solution to the $k$-SAT problem. We now need to check for which values of $\alpha$ this approximation holds. To do so, we write in the limit $n\longrightarrow \infty$
\begin{equation*}
\E[X^2]=2^nn\int_{0}^{1}d\zeta {n\choose\zeta n}f(\zeta)^m
\end{equation*}
By Stirling's formula, 
\begin{equation*}
{n\choose\zeta n}\sim \frac{1}{\sqrt{n}} \e^{nh(\zeta)}
\end{equation*}
where $h(\zeta)=-\zeta\ln(\zeta)-(1-\zeta)\ln(1-\zeta)$, so that 
\begin{equation*}
\E[X^2]=2^n\sqrt{n}\int_{0}^{1}d\zeta \e^{n\phi(\zeta)}
\end{equation*}
with $\phi(\zeta)=h(\zeta)+\alpha\ln{f(\zeta})$. Using the Laplace method, we can write:
\begin{equation*}
\E[X^2]\sim 2^n\sqrt n\sqrt{\frac{2\pi}{n\lvert\phi^{''}\rvert}} \e^{n\phi^{\text{max}}}
\end{equation*}
When $\phi^{\text{max}}=\phi(\frac{1}{2})$, corresponding to $\e^{\phi(\frac{1}{2})}=2(1-2^{-k})^{2\alpha}$, we recover that $\E[X^2]\sim 4^n(1-2^{-k})^{2m}=\E[X]^2$, consistently with a previous remark. For values of $\alpha$ where this holds, there exists a solution to the $k-SAT$ problem. But we know that can't be true for all values of $\alpha$, because we know there is a transition. We therefore need to compare the actual maximum of $\phi$ with $\phi^{\text{max}}$. But because $h$ is symmetric with respect to the $\zeta=\frac{1}{2}$ axis, and $\ln{f}$ is increasing, it is clear that whenever $\alpha>0$ also $\phi^{\text{max}}>\phi(\frac{1}{2}$), so that in the limit were $n$ goes to infinity, the second moment inequality (1) just tells us that $\P[X>0]\geq0$.

The way out is to consider another problem for which the function $f$ is also symmetric with respect to the $\zeta=\frac{1}{2}$ axis. Consider the NAE $k$-SAT problem, where we ask that at least one literal in each clause be false. It is clear that a truth assignment that satisfies an NAE $k$-SAT formula is also a solution to the corresponding $k$-SAT problem, so that a lower bound for $\alpha$ in NAE also holds in $k$-SAT. It turns out the function $f$ in NAE is given by 

\begin{equation*}
f(\zeta)=1-2^{1-k}+\zeta^k2^{1-k}+(1-\zeta)^k2^{1-k}
\end{equation*}
and therefore $\phi$ is symmetric around $\frac{1}{2}$. It is then straightforward to show that the maximum of $\phi$ is in $\frac{1}{2}$, up to some value of $\alpha=\frac{1}{2}2^k\ln2-O(1)$. We therefore have $\alpha_k\geq\alpha_k^{\text{NAE}}\geq\frac{1}{2}2^k\ln2-O(1)$, which is the lower bound we wanted. This lower bound can in fact be improved again to $2^k\ln2-O(1)$ by considering another random variable 
\begin{equation*}
X= \sum_{\sigma \text{ satis. } \phi} \eta^{\text{\# true literals}}
\end{equation*}
for some $\eta$ carefully chosen.

We conclude this section with some physical considerations on what the second moment inequality (\ref{eq_ineq1}) means. It is interesting to interpret this  inequality in terms of the planted ensemble. In the planted ensemble, we start by choosing at random a truth assignment $\sigma$, and then sample formulas $\phi$ such that $\sigma$ satisfies $\phi$. The expectation value of $X^2$ can be expressed in terms of the planted average in the following way:
 \begin{align*}
 \E[X^2]
 &=\sum_{\sigma,\tau} \P[\sigma,\tau\text{ sol.}]\\
 &=\sum_\sigma \P[\sigma\text{ sol.}]\ \sum_\tau \P[\tau\text{ sol. }\lvert\ \sigma\text{ sol.}]\\
 &=\E[X]\E[X\lvert\sigma]\\
 &=\E[X]\E_{\text{planted}}[X]
 \end{align*}
 so that equation (\ref{eq_ineq1}) can be rewritten 
 \begin{equation*}
 \P[X>0]\geq\frac{\E[X]}{\E_{\text{planted}}[X]}
 \end{equation*}
 By construction, $\E_{\text{planted}}[X]$ is bigger than $\E[X]$, because a formula constructed in the planted ensemble has more solutions on average than a completely random instance. But we see here that if the number of solutions is not too different, i.e. if the planted expectation is not too different from the exact expectation in the large $n$ limit, the second moment inequality tells us something new.

\section{Community detection}

\subsection{The stochastic block model}
It has been a general trend lately to consider everything as a network, either in physics, sociology, biology, finance\ldots  The question that naturally arises is whether this is justified, that is if seeing a system as a network actually allows to answer questions about the system.

In this section we consider the problem of community detection, i.e. the problem of identifying groups of nodes in a graph that share a common set of features.  
Community structure is called \emph{assortative} if each group has a larger connectivity between its members than with other communities. The opposite situation is called \emph{disassortative}. While trying to guess communities in a general setting, one doesn't know \textit{a priori} if they are associative or disassortative. 

To generate instances of this problem, we consider the Stochastic Block Model (SBM). It is a generalization of the Erd\H{o}s-Reny\'i ensemble, with $k$ types of vertices. These types can be thought of as colors, groups, spins, and so on. We start by supposing that $k$ is known, and we encode the parameters of the model in a $k\times k$ affinity matrix $p$ where $p_{r,s}$ is the probability that a given node from group $r$ is connected with a given node from group $s$. Denoting the type of node $i$ as $t_i$, we then construct the graph with the rule 
\begin{equation*}
\P[(i,j)\in E]=p_{t_{i},t_{j}} \, . 
\end{equation*}
Our goal is, given the graph $G$, to find the types of the nodes. 

Let us begin by rewriting the problem as a statistical physics problem. The probability of a given graph $G$ conditioned on the types of the nodes $t$ and the affinity between communities $p$ is given by
\begin{equation*}
\P[G\lvert t,p]=\underset{(i,j)\in E}{\prod}p_{t_i,t_j}\underset{(i,j)\notin E}{\prod}(1-p_{t_i,t_j})
\end{equation*}

In the problem we are considering, $t$ is unknown---it is what we are looking for. But we might know a priori probabilities $q_r$ for $r\in \{1,..k\}$ that a given node has type $r$. We then have 
\begin{align*}
\P[G\lvert p,q]&=\sum_t  \P[G\lvert t,p] \,\P[t\lvert q]\\
&=\sum_t underset{(i,j)\in E}{\prod}p_{t_i,t_j}\underset{(i,j)\notin E}{\prod}(1-p_{t_i,t_j})\ \prod_i q_{t_i}\\
&=\underset{t}{\sum} \e^{-H(t)} \, . 
\end{align*}
This is the partition function of a physical system at inverse temperature $\beta=1$, with Hamiltonian
\begin{equation}
  H(t)=-\underset{i\sim j}{\sum}\ln{p_{t_i,t_j}}-\underset{i\nsim j}{\sum}\ln{(1-p_{t_i,t_j})}-\underset{i}{\sum}\ln q_i
  \label{eq_ham}
\end{equation}
corresponding to a generalized Potts model with coupling constants $J_{ij}=\ln{p_{t_i,t_j}}$ and external fields $h_i=\ln{q_i}$. Note that it includes interactions between non-neighboring sites.  In the sparse case $\big(p_{rs}=\frac{c_{rs}}{n}\big)$, the coupling between two non-neighboring sites $i\nsim j$ is of order $\frac{1}{n}$.  However, we cannot simply get rid of these interactions, since the sum contains $O(n^2)$ such terms. 

The Boltzmann distribution over assignments of types $t$ is given, using Bayes' rule, by
\begin{align*}
\P[t\lvert G,p,q]&=\frac{\P[G,t\lvert p,q]}{\P[G\lvert p,q]} \\
&= \frac{\P[G\lvert t,p]\,\P[t\lvert q]}{\P[G\lvert p,q]} \\
&= \frac{\e^{-H(t)}}{\P[G\lvert p,q]}
\end{align*}
Now suppose we want to find the parameters $p$, $q$ that maximize $\mathcal{Z}=\P[G\lvert p,q]$; 
that is, we want to maximize the total probability of the network, summed over all type assignments.  
We can relate these optimal values of the parameters to thermodynamic quantities in the following way, 
\begin{align*}
\frac{\partial P}{\partial p_{rs}}&=\sum_t  \frac{\partial }{\partial p_{rs}} \e^{-H(t)}\\
&=-\underset{t}{\sum} \e^{-H(t)}\frac{\partial H(t) }{\partial p_{rs}} \\
&=-\underset{t}{\sum} \e^{-H(t)}\Big(-\underset{\underset{t_j=s}{\underset{t_i=r}{i\sim j}}}{\sum}\ \frac{1}{p_{rs}}+\underset{\underset{t_j=s}{\underset{t_i=r}{i\nsim j}}}{\sum}\ \frac{1}{1-p_{rs}}\Big) \, , 
\end{align*}
where $m_{rs}$ denotes the number of edges between group $r$ and group $s$, and $n_r$ denotes the number of nodes in group $r$. In a particular instance of the problem, we have 
\begin{align}
\frac{\partial P}{\partial p_{rs}}&=-\sum_t  \e^{-H(t)}\Big(-\frac{m_{rs}}{p_{rs}}+\frac{n_rn_s-m_{rs}}{1-p_{rs}}\Big) 
\nonumber \\
&\propto -\frac{\langle m_{rs}\rangle}{p_{rs}}+\frac{\langle n_rn_s\rangle-\langle m_{rs}\rangle}{1-p_{rs}}
\label{eq:diff-p}
\end{align}
where the brackets denote the average over the Boltzmann distribution. 

We will assume $\langle n_r n_s\rangle=\langle n_r\rangle\langle n_s\rangle$; this holds, in particular, if both $n_r$ and $n_s$ are tightly concentrated with $O(\sqrt{n})$ fluctuations.  
In that case, \eqref{eq:diff-p} gives
\begin{align*}
p_{rs}=\frac{\langle m_{rs}\rangle}{\langle n_r\rangle\langle n_s\rangle} \, . 
\end{align*}
Similarly, by taking the derivative
of $P$ with respect to $q_r$, we find 
\begin{align*}
q_r=\frac{\langle n_r\rangle}{n} \, . 
\end{align*}
Thus if we can estimate the averages with respect to the Boltzmann distribution, we can learn the parameters using the following Expectation-Maximization algorithm:
\begin{itemize}
\item E step: Compute the averages $\langle m_{rs}\rangle$ and $\langle n_r\rangle$ using the current values of $p$ and $q$.
\item M step: Update $p$ and $q$ to their most likely values given $\langle m_{rs}\rangle$ and $\langle n_r\rangle$.
\end{itemize}
We iterate until we reach a fixed point.

The only question that remains is how to estimate averages with respect to the Boltzmann distribution. This can be done by doing a Monte Carlo (MC) simulation, 
also known as Gibbs sampling.  For instance, we can use the Heat Bath algorithm: at each step, a node in the graph is chosen and ``thermalized,'' which means that its group $t_i$ is sampled from the marginal distribution imposed by its neighbors. More precisely, we set $t_i=s$ with probability proportional to $q_s\underset{j\sim i}{\prod}p_{s,t_j}$.  It is straightforward that this algorithm verifies detailed balance with respect to the Boltzmann distribution (\ref{eq_ham}).  After convergence, it will sample configurations with the correct weights.  
However, in order to compute averages like $\langle m_{rs}\rangle$, we need many independent samples, which forces us to run the algorithm for a large multiple of the autocorrelation time.

We can do better using Belief Propagation (BP). The idea of BP is that vertices pass each others estimates of marginals until consistency (a fixed point) is achieved. More precisely, we write
\begin{equation*}
\mu_r^{i\rightarrow j}=\P[t_i=r \text{ if } j \text{ were absent}] \, . 
\end{equation*}
which is the cavity interpretation of the messages of BP.  
Our goal is to estimate the one-node and the two-node marginals:
\begin{align*}
\mu_r^{i}&= \P[t_i=r]\\
\mu_{rs}^{ij}&= \P[t_i=r,\ t_j=s]
\end{align*}
BP estimates of these marginals can be expressed in terms of the messages.  In particular, 
\begin{align*}
\mu_{rs}^{ij} \propto \mu_{r}^{i\rightarrow j} \mu_{s}^{j\rightarrow i} 
\left\{
    \begin{array}{ll}
        p_{rs} & \mbox{if } i\sim j \\
        1-p_{rs} & \mbox{otherwise}
    \end{array}
\right.
\end{align*}

The BP update rule is given by
\begin{equation}
\label{eq:bp-update}
\mu_r^{i\rightarrow j}=\frac{1}{\mathcal{Z}^{i\rightarrow j}} 
q_r 
\prod_{\substack{k \sim i \\ k \ne j}} \sum_s \mu_s^{k\rightarrow i} p_{rs} 
\prod_{\substack{k \not\sim i \\ k \ne j}}
 \sum_s \mu_s^{k\rightarrow i} (1-p_{rs})
\end{equation}
As usual in BP, this expression assumes that nodes other than $i$ are independent conditioned on $t_i$. This is only approximately true if $G$ is not a tree; 
we believe that it is approximately true when $G$ is locally treelike, as in graphs generated by the sparse stochastic block model. Even in real networks, it works surprisingly well.

Note that~\eqref{eq:bp-update} actually takes place on a fully connected graph, because of the interaction between non-neighboring sites. This yields $O(n^2)$ calculations at each update. To recover sparsity in this formula, we will assume that site $i$ only feels the mean field of its non-neighboring sites, that is $\mu_r^{k\rightarrow i}=\mu_r^k$ for all $k\nsim i$. With this assumption, one iteration of BP requires only $O(m)$ computations, where $m$ is the number of edges: for sparse graphs, we have $m=O(n)$ rather than $O(n^2)$.  This makes the algorithm far more scalable.



Once the parameters  $p,q$ are set correctly, we can run MC or BP to determine the most likely assignment of group types. But we get a much finer sense of what's really going on by taking a look at the whole distribution of group assignments.  Zachary's Karate Club provides a cautionary tale about the risks of the procedure we just described. Wayne Zachary collected relationship data in a university karate club composed of 34 people in 1977. Due to an argument between the president of the club and the instructor, the club split up in two groups. Some followed the president, some others followed the instructor. Zachary asked whether it is possible, from the friendship network he collected before the split-up, to retrodict the composition  of the two groups. This is easy to do except for one node, who was closer to the president, but ended in the instructor's group because he had a black belt exam three weeks later and didn't want to change instructor!

This story tells us that the graph cannot contain all the information relevant to assigning communities to nodes. On the other hand, by looking at the distribution of group assignments $t$, we can get a sense of how tightly each node is linked to its community. For instance, we could have computed that the president, the instructor, and their closest friends had a $99\%$ chance of ending up in their respective groups, while some other nodes only had a $60\%$ chance to end up in a given group. In this sense, determining the assignment $t$ is rather a marginalization than a maximization problem.

A somewhat similar problem arises at the level of determining the parameters $p$ and $q$.  In the network literature, many authors vary the parameters $p,q$ in order to minimize the ground state energy.  That is, they minimize 
\begin{equation*}
\P[t\lvert G,p,q]=\frac{\e^{-H(t)}}{\mathcal{Z}} \, .
\end{equation*}
The problem with this approach is that by varying the parameters $p,q$, one can reach artificially low ground state energies. \cite{zdeborova2010conjecture} showed that in a random 3-regular graph, one can always find a partition of the nodes in two groups such that only $11\%$ of the edges cross from one group to the other. One might think that such a partition shows communities in the graph, although it was generated completely at random. The point is that there are exponentially many ways to partition the nodes in two groups, and it is hardly astonishing that there should exist one with few edges between the two groups: 
but this is really just overfitting, fitting the random noise in the graph rather than finding statistically significant communities.  Instead of looking at the ground state energy, one should focus on the free energy $F=-\ln{\mathcal{Z}}=-\ln{P[G\lvert p,g]}$.  This is another reason we prefer BP to MC. In MC, computing the entropy is tricky because it requires us to integrate over different temperatures, and running MC for a long time at each temperature. 

To see how BP provides us with an estimate of the free energy, we transform the problem of evaluating $F$ into a variational problem. For simplicity we write $\P[\cdot]$ for $\P[\,\cdot \mid p,q]$. 
Let $\Q$ be an arbitrary probability distribution over the possible assignments $t$.  The free energy is
\begin{align*}
-F = -\ln \mathcal{Z} &=\ln{\P[G]}\\
&=\ln{\sum_t  \P[G,t]}\\
&=\ln{\sum_t  \Q(t)\frac{\P[G,t]}{\Q(t)}}\\
&=\ln{\E_{t\sim \Q(t)} \frac{\P[G,t]}{\Q(t)}}
\end{align*}
Jensen's inequality (i.e., the concavity of the logarithm) gives $\ln \E X \ge \E \ln X$.  Then 
\begin{align*}
-F&\geq \E_{t\sim \Q(t)} \ln{\frac{\P[G,t]}{\Q(t)}}\\
&\geq \E_{t\sim \Q(t)} \ln{\P(t)}-\sum_t  \Q(t)\ln{\Q(t)}
\end{align*}
so that
\begin{align*}
F\leq E_\Q-S(\Q)
\end{align*}
where $E_\Q$ is the average energy if a configuration $t$ has probability $\Q(t)$ instead of the Boltzmann probability $\P$, and $S(\Q)$ is the entropy of the distribution $\Q$. Furthermore, this inequality is saturated if and only if $\P=\Q$. 

This allows us to approximate the free energy by minimizing the Gibbs free energy $E_\Q-S(\Q)$. To make the variational problem tractable, we can constrain $\Q$ within a family probability distributions with a small (i.e., polynomial) number of parameters.  
A popular choice is the mean-field assumption, in which correlations between different sites are neglected.  
This assumes that $\Q$ is simply a product distribution, 
\begin{equation*}
\Q(t)=\prod_{i} \mu_{t_i}^i \, .
\end{equation*}
BP does better than this rough assumption as it considers two-node correlations, in addition to single-site marginals.  In fact, as was shown by \cite{yedidia2000generalized}, BP fixed points are the stationary points of the free energy within the family where $\Q$ is of the form 
\begin{equation*}
\Q(t)=\frac{\prod_{i\sim j}  {\mu_{t_i,t_j}^{i,j}}}{\prod_i \mu_{t_i}^{d_i-1}}
\end{equation*}

Note that this form isn't even a distribution unless the underlying graph $G$ is a tree. Plugging this form into the Gibbs free energy in general therefore only gives an approximation of the Helmholtz free energy, called the Bethe free energy. 

\begin{figure}[]
\centering
   \includegraphics[scale=1]{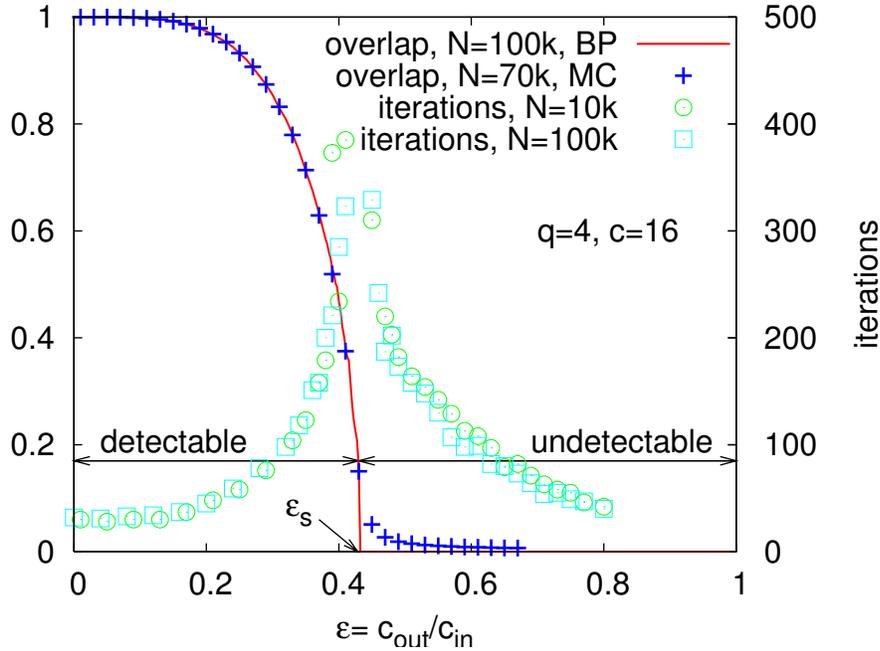}
   \caption{Second order phase transition for $k\leq4$, from \cite{zdebo11}}
  
   \label{fig_ralentBP}
\end{figure}

Let us now see how BP performs on a concrete example.  Again letting $k$ denote the number of groups, we generate a graph using the following parameters of the Stochastic Block Model:
\begin{align*}
q_r&=\frac{1}{k}\\
p_{rs}&=\frac{c_{rs}}{n}\\
c_{rs}&=\left\{
    \begin{array}{ll}
        \cin & \mbox{if } r=s \\
        \cout & \mbox{if } r\neq s
    \end{array}
\right.
\end{align*}
The average connectivity is 
\begin{equation*}
c=\frac{1}{k}\cin+\frac{k-1}{k}\cout \, . 
\end{equation*}
Given the graph, the goal is to infer an assignment of the groups $t$ that maximizes the overlap with the true values. Figure~\ref{fig_ralentBP} shows the running time of BP as a function of the ratio $\cout/\cin$. We notice that it has hardly any dependence on the number of nodes $n$, and that at some particular value $c^*$ a critical slowdown appears, typical of second-order phase transitions. 

It is in fact possible to evaluate analytically at which value of the parameters BP begins to fail to detect the communities. All the nodes have the same average degree, so that the uniform distribution over all nodes is a fixed point of the BP equations. Let us study the linear stability of this fixed point. To do so, we introduce a small perturbation in the following way:
\begin{equation*}
\mu_s^{i\rightarrow j}=\frac{1}{k}+\epsilon_s^{k\rightarrow i}
\end{equation*}
The linearization of the BP equations then takes the form:
\begin{equation*}
\epsilon^{i\rightarrow j} =
\sum_{\substack{k \sim i \\ k \ne j}} 
T\epsilon^{k\rightarrow i}
\end{equation*}
where $T$ is a $k\times k$ matrix with components 
\[
T_{rs}=\frac{1}{k} \left( \frac{c_{rs}}{c}-1 \right) \, . 
\] 
The eigenvalues of this matrix are $0$ and 
\[
\lambda=\frac{\cin-\cout}{kc} \, . 
\]
If $\lambda$ is too small, the uniform distribution is a stable fixed point of BP, and we won't learn anything by running it: 
it will simply conclude that every node is equally likely to be in every group. 

More precisely, if we assume that distant nodes are independent, then the perturbation $\epsilon$ gets multiplied by a factor $( \lambda \sqrt{c})^d$ at level $d$ of the tree. The stability condition of the uniform distribution is therefore $\lambda \sqrt{c} < 1$, and the transition happens at
\begin{equation*}
\cin-\cout=k\sqrt{c}.
\end{equation*}
Thus BP fails if $\lambda \sqrt{c} < 1$, labeling the nodes no better than chance.  Our claim is that any other algorithm will also fail in this region: the randomness in the block model ``washes away'' the information of the underlying types.  

Indeed, after this argument was presented by \cite{zdebo11}, It was shown rigorously and independently by \cite{mossel2013proof} and \cite{massoulie2013community} in the case of $k=2$ groups of equal size that if $\lambda \sqrt{c} < 1$, the marginal distributions of the nodes approach the uniform distribution.  In fact, the ensemble of graphs generated by the Stochastic Block Model is contiguous to the Erd\H{o}s-Reny\'i ensemble, meaning that one graph is not enough to distinguish the two ensembles: there is no statistical test that determines, with probability approaching $1$, whether communities exist or not.  Conversely, if $\lambda \sqrt{c} > 1$, then a BP-like algorithm can indeed label the nodes better than chance.  

\begin{figure}[]
\centering
   \includegraphics[scale=1]{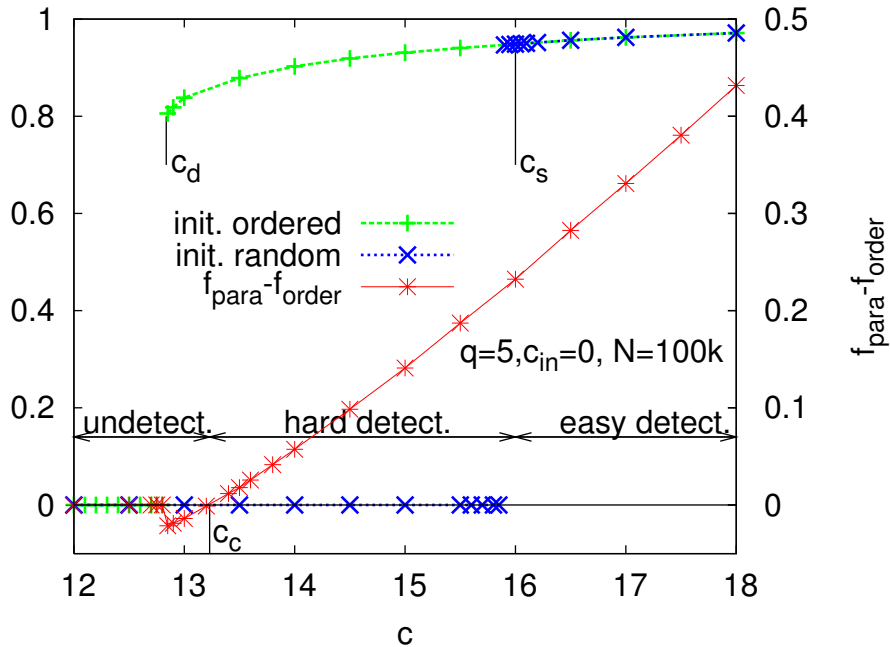}
   \caption{First order phase transition for $k>4$, from \cite{zdebo11}}
         \label{fig:firstOrder}
\end{figure}

One may then ask what happens if we have more than two groups. It turns out that for $k>4$, the situation is different, and the transition is first order (see figure \ref{fig:firstOrder}). The purple curve indicates the probability of detection when BP starts from a random
initial condition. This is called robust reconstruction, but it fails at an ``easy/hard transition'' well above the detectability threshold $c_d$.  On the other hand, if we give BP a hint, starting from a configuration not too far from the true assignment, then it achieves the detectability threshold (green curve). In between the detectability transition and the easy/hard transition, 
there are two fixed points of BP: the uniform one, and an accurate one.  The Bethe free energy of the accurate was is lower, so we would choose it if we knew it; however, its basin of attraction is exponentially small.  Thus we have a ``hard but detectable'' regime where detection is information-theoretically possible, but (we believe) exponentially hard. 


\subsection{Spectral methods}

We now turn to another 
family of algorithms for
community detection. These are called spectral methods, and aim at inferring types of nodes by diagonalizing a matrix which encodes the graph we are given. Popular choices for this matrix include the adjacency matrix $A$, the random walk matrix $D^{-1}A$ (where $D$ is the diagonal matrix where $D_{ii}$ is the degree $d_i$ of node $i$) the Laplacian $D-A$, and the symmetrically normalized Laplacian $D^{-1/2}(D-A)D^{-1/2}$. 

Generically, the largest eigenvalues of these matrices will be correlated with the community structure---except for the Laplacian, where one has to look at the smallest positive eigenvalues. More precisely, the largest eigenvector 
depends on the node degree or other notions of ``centrality'' (see for instance \cite{newman2010networks}), while (for $k=2$ groups) the second eigenvector will be positive in one group and negative in the other.
However, for the matrices listed above, this program will work up to a value of $\cin-\cout$ 
slightly above the BP threshold. Can we understand why?

The problem is that the spectrum of sparse random matrices is more difficult to study than that of dense matrices. If a random matrix is dense enough, with the average degree 
growing at least like $\log n$, then the spectrum can be modeled as a discrete part related to the community structure, and a continuous ``bulk'' that follows Wigner's semicircle law, lying in the interval $[-2\sqrt{c}, 2\sqrt{c}]$ (see \cite{nadakuditi2012graph}).  However, if the average degree is $O(1)$, the distribution of eigenvalues differs from Wigner's semi-circle law. In particular, tails appear at the edges of the semi-circle due to the high-degree vertices, which might drown the informative eigenvalue. 

To make this argument more precise, it is easy to see that the adjacency matrix $A$ (for example) has an eigenvalue $\lambda$ such that 
\begin{equation*}
|\lambda| \ge \sqrt{\dmax}
\end{equation*}
where $\dmax$ is the largest degree of a node in the graph.  To see this, consider a node $i$ with degree $\dmax$, and let $e_{i}$ be the vector where is $1$ at $i$ and zero elsewhere.  Note that $(A^k)_{ij}$ is the number of paths from $i$ to $j$ in the graph of exactly $k$ steps.  Since there are $d_i = \dmax$ ways to leave $i$ and return with two steps, we have $A^2_{i,i} = \dmax$.  Taking inner products, we have
\[
\langle e_i, A^2 e_i \rangle = \dmax \, . 
\]
Since $A^2$ is symmetric and $e_i$ can be orthogonally decomposed as a linear combination of eigenvectors, it follows that $A^2$ has an eigenvector at least $\dmax$, and $A$ has an eigenvector which is at least $\sqrt{\dmax}$ in absolute value.

In the Erd\H{o}s-Reny\'i ensemble we have $\dmax = O(\log n / \log\log n)$.  Thus for sufficiently large $n$, $\lvert\lambda\rvert$ becomes bigger than the edge of the Wigner semicircle.  
As a consequence, spectral methods based on $A$ or $L$ will tend to find localized eigenvectors, clustered around high-degree vertices, as opposed to finding those correlated with the community structure.  The normalized versions of $A$ and $L$ have analogous problems, where they get stuck in the trees of size $O(\log n)$ that dangle from the giant component.  Thus, in the sparse case where the average degree is $O(1)$, all these methods fail significantly above the detectability threshold.

\begin{figure}[]

\centering
   \includegraphics[scale=0.7]{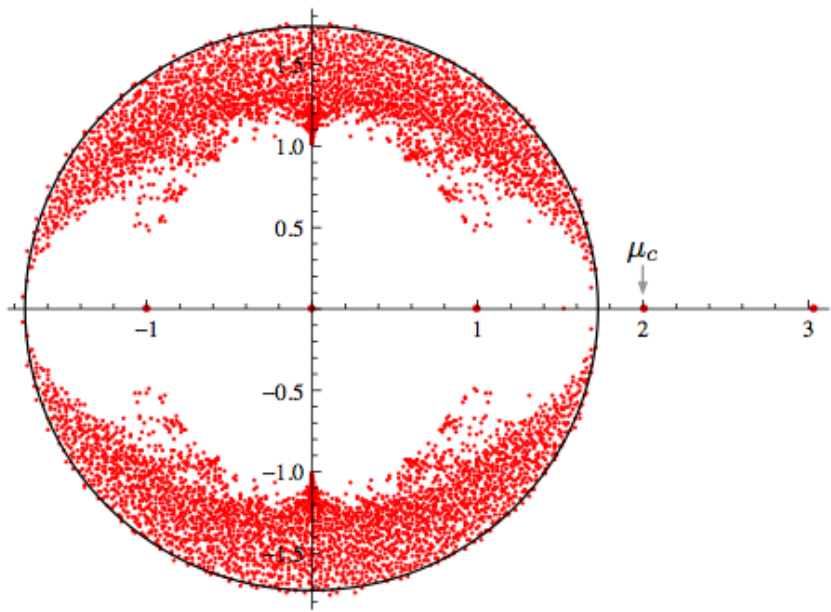}
   \caption{The eigenvalues of the non-backtracking matrix, from \cite{zdebo13}}
   \label{fig:B}
\end{figure}

We therefore see that if we want to achieve optimal community detection by spectral methods, we need to design a matrix that forbids going back and forth on the same edge.  
The simplest way to do this is to define a matrix $B$ that describes a non-backtracking walk on the edges of the network,
\begin{equation*}
B^{\text{non-backtracking}}_{i\rightarrow j,k\rightarrow \ell}=\delta_{jk}(1-\delta_{i\ell})
\end{equation*}
\cite{zdebo13} conjectured that even in the sparse case, all of $B$'s eigenvalues are inside a circle of radius $\sqrt{c}$ in the complex plane, except for the leading eigenvalue and the ones correlated with community structure.  This conjecture is supported by extensive numerical experiments (see Fig. \ref{fig:B}).  
Moreover, $B$ describes to first order how messages propagate in BP, and in particular the linear stability of the uniform fixed point discussed above.  Thus, if this conjecture is true, spectral clustering with $B$ succeeds all the way down to the detectability transition for $k=2$, and to the easy/hard transition for larger $k$.

As a concluding remark, we should keep in mind that this discussion assumed that the network is generated by the stochastic block model.  In fact, for many networks this model is inaccurate; for instance, it doesn't produce heavy-tailed degree distributions, which are common in real networks.  The good news is that many of these techniques, including belief propagation, the Bethe free energy, and so on, are easy to extend to more elaborate models, such as the degree-corrected block model of \cite{karrer2011stochastic}. 


\section{Appendix}

\subsection{Definition of Perfect Matching}
The perfect matching problem consists in finding a set of matching between nodes such that it is compatible with the structure of the network. Let's consider the graph $G=(V,E)$ of vertices and edges. It aims to find a set of disjoint edges 
that covers as many nodes as possible, giving each of those nodes a unique partner. A perfect matching is a matching where all the nodes are covered. 

\subsection{Definition of Max Flow}
Consider a directed graph $G=(V,E)$, with a source node $s$ and a sink node $t$. Each edge $e$ has a given capacity $c(e)$ which is a non-negative integer.   
A flow assigns a number $f(e)$ to each edge so that the total flow in and out of each node is conserved, except for the source and sink.  We require that $0 \le f(e) \le c(e)$, and we seek to maximize the total flow, i.e., the sum of $f(e)$ over all the outgoing edges of $s$ (or the incoming edges of $t$).

\subsection{Definition of k-SAT}
An instance of $k$-SAT is a formula involving $n$ Boolean variables $x_1, \ldots, x_n$.  There are many variations such as NAESAT, $1$-in-$k$ SAT, and so on, but in the standard version of $k$-SAT the formula is in ``conjunctive normal form.''  That is, it is the AND of a set of clauses; each clause is the OR of a set of literals; and each literal is either a variable $x_i$ or its negation $\overline{x}_i$.  Thus a truth assignment, i.e., an assignment of true/false values to the variables, satisfies a clause if its makes at least one of its literals true, and it satisfies a formula if and only if it satisfies all its clauses.  A formula is satisfiable if a satisfying assignment exists.  

\bibliographystyle{plain}
\bibliography{references}

\end{document}